\documentclass[prc,twocolumn]{revtex4-2}

\usepackage{graphicx}
\usepackage{color}
\usepackage{amsmath,bm}
\usepackage{amssymb}
\usepackage{amsthm}
\usepackage{mathrsfs}
\usepackage{array}
\usepackage{latexsym}
\usepackage{slashed}
\usepackage{hyperref}

\newcommand{\be}{\begin{equation}}
\newcommand{\ee}{\end{equation}}
\newcommand{\ba}{\begin{eqnarray}}
\newcommand{\ea}{\end{eqnarray}}
\newcommand{\ban}{\begin{eqnarray*}}
\newcommand{\ean}{\end{eqnarray*}}
\newcommand \nn {\nonumber}

\begin{document}

\title{QCD Phase Boundary and the Hadrochemical Horizon}
\author{Berndt M\"uller}
\affiliation{
Department of Physics, Duke University, Durham, North Carolina 27708, USA}
\affiliation{
Department of Physics Wright Laboratory, Yale University, New Haven, Connecticut 06511, USA}

\begin{abstract}
I review the physics of the phase boundary between hadronic matter and quark matter from several different points of view. These include thermodynamics, statistical physics, and chemical kinetics. In particular, the review focuses on the role of the chemical freeze-out line and its relation to the concept of valence-quark percolation. The review ends with some recollections of Jean Cleymans.
\end{abstract}

\maketitle

\section{The QCD Phase Boundary}

The phase diagram of strongly interacting (QCD) matter contains a low-energy density region, in which the mobile constituents are color singlets (hadrons), and a high-energy density domain, in which the mobile constituents are color non-singlet objects, including quarks, antiquarks, gluons and, at high net baryon number density and modest temperature, diquarks. The low-energy density matter is commonly called ``hadronic matter'' or  ``hadron gas'';  the matter at high-energy density is called ``quark-gluon plasma'' or ``quark matter''.  Upon finer inspection the diagram may be subdivided into a multitude of specific phases, including various types of color superconductors \cite{Alford:2007xm} and a possible phase that has been named ``quarkyonic'' matter \cite{Andronic:2009gj}.

The first attempt to determine the boundary between hadronic matter and quark matter was made by Hagedorn and Rafelski in the framework of the statistical bootstrap model, where they found a continuous first-order transition line in the QCD phase diagram \cite{Hagedorn:1980kb}. However, as numerical simulations of QCD on a lattice have shown, for physical values of the quark masses the two domains are not everywhere separated by a true thermodynamic phase boundary, i.e., a line characterized by singularities in the thermodynamic partition function \cite{Bazavov:2017dus,Borsanyi:2020fev}. Yet it is widely believed that such singular boundary exists in some regions of the phase diagram. The most widely studied example of a thermodynamic singularity is a possible critical endpoint of a first-order line separating hadronic matter and quark matter, corresponding to the spontaneous breaking of chiral symmetry at moderate baryon number density and temperature. The first-order transition line may have a second critical endpoint induced by the axial U(1) anomaly at even lower temperature and larger net baryon density \cite{Hatsuda:2006ps}.

State-of-the-art lattice-QCD calculations show that the critical endpoint, if one exists, must be located in the 
region $\mu_B/T > 3.5$ of baryochemical potential, $\mu_B$, and temperature, $T$ \cite{Borsanyi:2021sxv}. Analytical calculations using the functional renormalization group method \cite{Fu:2019hdw} indicate that the critical endpoint (CEP) lies at $(T_\mathrm{CEP}, \mu_{B,\mathrm{CEP}}) \approx (107~\mathrm{MeV}, 635~\mathrm{MeV})$ corresponding to $\mu_{B,\mathrm{CEP}}/T_\mathrm{CEP} \approx 5.54$. A holographic model of QCD based on an Einstein--Maxwell-dilaton Lagrangian in the anti-de-Sitter AdS$_5$ five-dimensional space with parameters tuned to reproduce thermodynamic properties of QCD at low net baryon density \cite{Critelli:2017oub} predicts a similar position of the critical endpoint at $(T_\mathrm{CEP}, \mu_{B,\mathrm{CEP}}) \approx (89~\mathrm{MeV}, 724~\mathrm{MeV})$. At smaller values of $\mu_B/T$, where the transition between the two regimes is continuous, one can define 
a pseudocritical line as the location of the maximum of the chiral susceptibility for a fixed ratio $\mu_B/T$. The value of the pseudocritical temperature at \mbox{$\mu_B = 0$ is $T_c = 156.5 \pm 1.5$~MeV \cite{HotQCD:2018pds}.}

The phase boundaries defined by the properties of thermodynamic functions are not easily mapped onto experimentally accessible observables. In principle, the internal energy and the pressure enter into the hydrodynamic equations of motion via the equation of state but, in practice, the contributions of singular terms to the equation of state are small and are hardly visible in the expansion dynamics of a quark-gluon plasma fireball. This would be different if the phase boundary would correspond to a strong first-order phase transition with a large latent heat, but such a scenario is not realized in QCD matter. This motivates the consideration of other criteria for a phase boundary. One can distinguish three types of such criteria:
\begin{enumerate}
\item 
Thermodynamic phase boundary.
This is the definition  discussed above, where the phase boundary is defined as the locus of singularities in the thermodynamic functions, or as the vicinity of such singularities such as in QCD, where singular behavior only occurs in the two-flavor chiral limit.
\item 
Statistical phase boundary. 
In this case, the boundary is defined as the locus where certain statistical properties of the matter exhibit a singular behavior. Examples are critical fluctuations at or near a critical point (which also satisfies the thermodynamic criterion) or a percolation threshold. 
\item 
Kinetic phase boundary.
Such boundaries exist in dynamical scenarios when kinetic processes that slow down during a cooling or expansion process become much slower than the characteristic cooling or expansion rate and ``freeze out''. An example is the recombination of hydrogen atoms in the early universe, when the universe became transparent to the photons of the blackbody radiation. A more apt term for this type of boundary would be the photon horizon or optical horizon of the universe, as it is impossible to observe phenomena before that boundary via photons. In the case of the strongly interacting matter, the relevant reaction is the exchange of flavor quantum numbers, heavier than 
$u$ and $d$ quarks,  among hadrons. Here, this boundary 
is called the  ``hadrochemical horizon''; more commonly it is called the ``chemical freeze-out line''.
\end{enumerate}

\section{Valence-Quark Percolation}
\label{sect2} 

The na{\"i}ve concept of the deconfinement transition in QCD is that it separates the ``normal'' phase at low temperature and net baryon density, where quarks are confined to hadrons, from the high-energy density regions of the QCD phase diagram, in which quarks can exist as isolated excitations. In spite of its intuitive allure, this concept is obscured by the mechanism of quark-pair production, which permits a quark to move around easily by dressing itself with a light antiquark. Quark confinement has a rigorous definition in a world in which only gluons exist, i.\ e., in the pure SU(3) gauge theory, which can be thought of as the hypothetical limit of QCD with very large quark masses when pair production is energetically disfavored. In that idealized limit, quark confinement can be unambiguously identified by the vanishing of the expectation value of the Polyakov loop. In the presence of light dynamical quarks this measure never vanishes, and no rigorous definition of what one means by ``quark confinement'' in a world with light quarks has been 
found. 

This insight suggests that it may be more rewarding to look at quark confinement from a kinetic point of view rather 
than an energetic viewpoint. The transport of an individual quark, i.\ e., a quark identified by its distinct flavor 
from one location to another in the confined phase requires the transport of a hadron that contains this quark as a constituent. In the case of a strange quark, the least massive vehicle of transport is a kaon, which weighs approximately 0.5 GeV/c. At low temperature ($T < 100$ MeV) kaons are rare excitations in the hadronic gas; this reduces the rate of exchange of strange quarks between hadrons. This argument obviously does not apply to $u$ and $d$ quarks as they can be exchanged as constituents of pions, which provide for a long-range---on the nuclear scale---exchange mechanism. 
The exchange of a strange quark between hadrons thus predominantly occurs via direct exchange of two quarks---one from each hadron---when the quark cores of the two hadrons come (nearly) into contact. At low temperature or low net baryon density such close encounters among two hadrons are rare, and they are even rarer for close encounters between three or more hadrons.  Long-range quark transport, although  possible, is, thus, greatly impeded. In this picture, one can think of quark exchange among hadrons as an effective kinetic mechanism for quark transport.

The rate of close encounters of multiple hadrons obviously grows rapidly with increasing density, either by excitations of more hadrons at higher temperature or owing to increased net baryon density at low temperature. At sufficiently high hadron density, the quark cores of all hadrons will be in contact with another hadron core most of the time. There must exist some critical hadron density at which it is possible to find a chain of pairwise touching hadrons bridging across an arbitrary large distance; this critical density is commonly called the ``percolation threshold''. In this picture, quark deconfinement corresponds to a percolation transition and the QCD phase boundary is identified as a percolation boundary \cite{Magas:2003wi}.

This concept of valence-quark percolationis different from the concept of color string percolation~\cite{Armesto:1996kt,Braun:1997ch},  which has been invoked to describe the formation of a deconfined QCD plasma in the initial stage of a nuclear collision. In the color string percolation model the percolation criterion applies to the area density of color strings or flux tubes that are formed immediately after the collision of two nuclei; the model is closely related to the glasma model \cite{Kovner:1995ja,McLerran:2007zzc} for the initial collision stage. The concept of valence-quark percolation, in contrast, applies to the late stage of the collision when the quark-gluon plasma disassembles into hadrons. 

Since hadrons have no well defined surfaces, the valence-quark  percolation picture contains some level of ambiguity. For example, one could model hadrons in the spirit of the (Massachusetts Institute of Technology)  MIT-Bag Model \cite{Chodos:1974pn} either as hard spheres or as fuzzy spheres that can overlap. Alternatively, one could model hadrons as valence-quark  bags surrounded by meson clouds along the lines of the Cloudy Bag model \cite{Thomas:1981vc}. For a thermal mass distribution of hadrons, the excluded volume in the hard sphere model might be assumed to be proportional to the mass of the hadrons as in the MIT-Bag Model: $M = 4{\cal B}V$ with the bag constant ${\cal B} \approx (145~\mathrm{MeV})^4$and $V$ denoting the bag volume. Studies of the percolation threshold for various three-dimensional hard sphere models can be found elsewhere. \cite{Stauffer:2018,Grimaldi:2007,Desmond:2014}. The occupied volume fraction, $\xi$, where a volume spanning cluster appears, lies typically in the range,  $\xi_{\rm cl}  \approx$ 0.6--0.75. For reference, the volume fraction occupied by closely packed (CP) uniform hard spheres is: $\xi_\mathrm{CP} = \pi/(3\sqrt{2}) \approx 0.74$.

\section{Chemical Freeze-Out}

The quest for an experimental determination of the hadronic gas--quark-gluon plasma boundary is as old as the search for the quark-gluon plasma itself. It was already early recognized that hadron ratios, especially those of hadrons containing strange quarks, carry information about the temperature and baryochemical potential prevalent in the hadronic gas when it disassembles at the end of a heavy-ion collision \cite{Cleymans:1992zc,Suhonen:1993zy,Letessier:1994cn}. If these hadrons freeze out soon after the transition from quark-gluon plasma to the hadronic phase, the parameters deduced from thermal fits to the measured hadron yields would even be proxies for the phase boundary. 

The premise that hadrochemical abundances freeze out shortly after hadronization was first substantiated by Koch, M\"uller, and Rafelski \cite{Koch:1986hf,Koch:1986ud}, who studied the evolution of strange hadron abundances in a hot hadronic gas and concluded that they could not be appreciably changed by hadronic reactions at temperatures below the critical temperature: $T < T_c \approx 160~\mathrm{MeV}$, where a hadronic gas exists. A similar study by Braun--Munzinger, Stachel, and Wetterich \cite{Braun-Munzinger:2003htr} reached the same conclusion: The experimentally determined chemical freeze-out temperature of strange baryons is a good proxy for the phase transition temperature.

An early version of a QCD phase diagram with a (very sparse) set data points from  SPS  (Super Proton Synchrotron) and  AGS (Alternating Gradient Synchrotron) experiments can be found in the review by Harris and M\"uller \cite{Harris:1996zx} (see Figure~4 there).  Using a larger set of data points from the SPS, AGS and SIS (Schwerionensynchrotron, Heavy Ion Synchrotron),  Cleymans and Redlich \cite{Cleymans:1998fq} observed that the freeze-out line, which we above called the hadrochemical horizon, corresponds to a fixed energy per emitted particle, $\langle E \rangle / \langle N \rangle \approx 1~\mathrm{GeV}$,  where the brackets denote sample averaging.  This contrasts with the condition for the phase boundary which, in the statistical bootstrap model with excluded volume of Hagedorn and Rafelski \cite{Hagedorn:1980kb}, corresponds to a fixed critical  energy density, $\varepsilon_{\rm c} = 4{\cal B}$. The value $\varepsilon_{\rm c} = 0.23~\mathrm{GeV/fm}^3$ is close to the value of the energy density at the minimum of the speed of sound in lattice-QCD calculations, \mbox{$\varepsilon_\mathrm{min} = 0.20~\mathrm{GeV/fm}^3$} \cite{Borsanyi:2010cj}.This was a surprising result, which has since been confirmed by even richer data sets but still lacks a simple explanation. One aspect that made the result surprising is that the particle content varies greatly along the freeze-out line, being meson-dominated at low $\mu_B$ but dominated by baryons at high $\mu_B$.  The observation showed, however, that the chemical freeze-out line cannot generally be regarded as a proxy for the phase boundary, even if it closely tracks  it at low and moderate baryochemical potential.

\begin{figure}[htb]
	\includegraphics[width=0.95\linewidth]{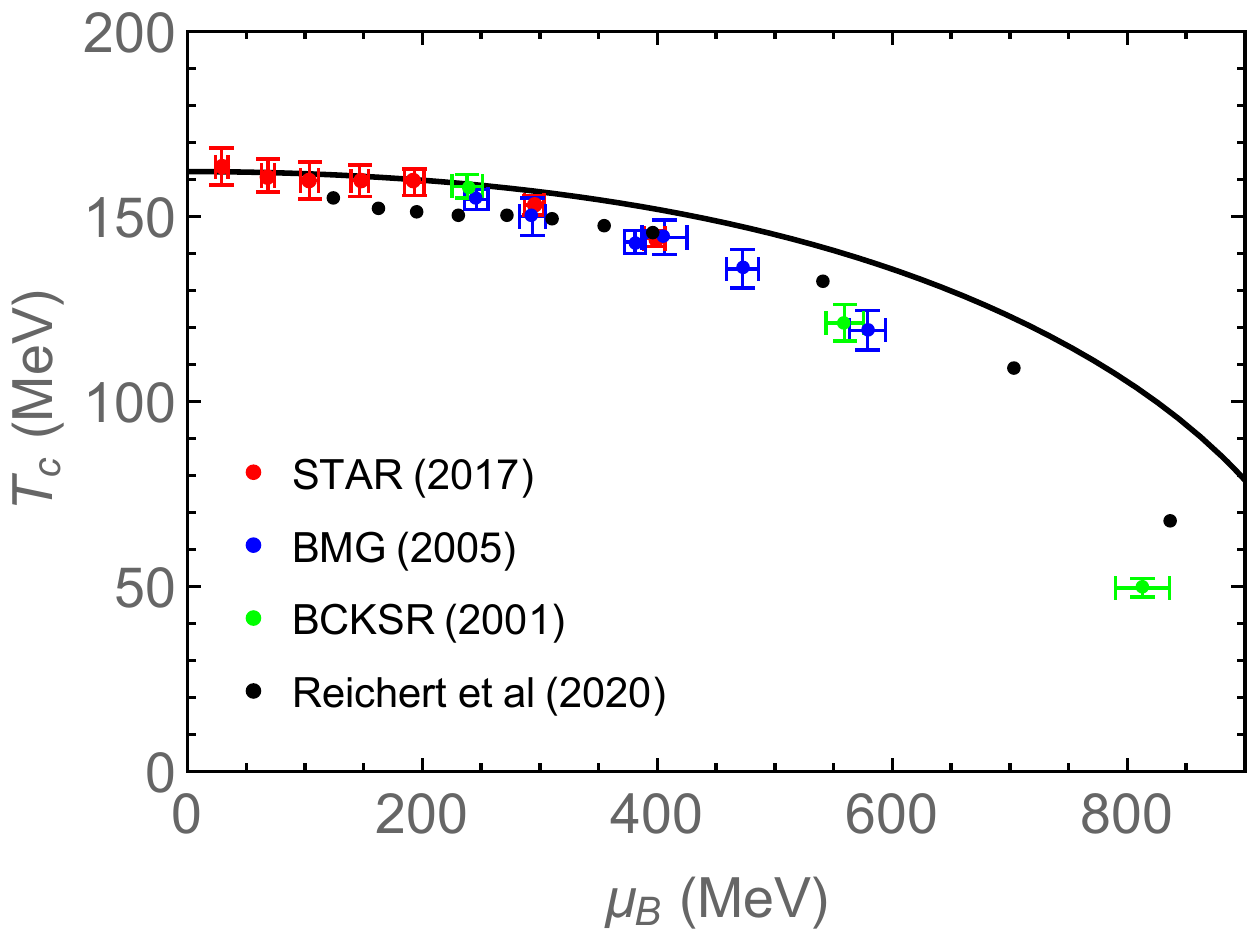} 
	\caption{Chemical potential and temperature at chemical freeze-out, determined by thermal model fits to strange 		hadron yield ratios for data from AGS, RHIC, and LHC  experiments  
	\cite{STAR:2017sal,Andronic:2005yp,Flor:2020fdw}. 
	The STAR experimentdata (red) are for 0--10\% central collisions. 
	The black dots show the freeze-out parameters obtained from simulations using a hadronic transport model 			\cite{Reichert:2020yhx}.
	Blue (BMG) and green (BCKSR) dots are calculations by Becattini {\em et al.} from Refs. \cite{Becattini:2005xt} and 		\cite{Becattini:2000jw}, respectively (denoted after the authors' names). 
	The solid line shows the hadron gas phase boundary obtained in the statistical bootstrap model with excluded volume 
	(see Section~\ref{sect4} below).}
	\label{fig:Chem}
\end{figure}

An extensive set of hadrochemical freeze-out data was obtained in the RHIC (Relativistic Heavy Ion Collider) beam energy scan (BES) \cite{STAR:2017sal}, which cover the range $\mu_B < 400~\mathrm{MeV}$. These will soon be complemented by the higher statistic data from the BES-II at RHIC, which extends at least up to $\mu_B \approx 700~\mathrm{MeV}$. Comprehensive thermal fits were performed by Becattini {et al.} \cite{Becattini:2005xt} for data from SPS, AGS, and SIS, and by Andronic {et al.} \cite{Andronic:2005yp} for both, midrapidity and $4\pi$ data from AGS, SPS, and RHIC. Separate fits to STAR and ALICE experiments data for strange and nonstrange hadrons was recently published by Flor {et al.} \cite{Flor:2020fdw}. The results for some of these chemical freeze-out fits are shown in Figure~\ref{fig:Chem}.
The analyses compare various measured (``exp'') 
particle yields per unit rapidity, $dN_i^\mathrm{(exp)}/dy$, with the thermal predictions 
(``th'') 
for these yields:
\be
\frac{dN_i^\mathrm{(th)}}{dy} = \frac{dV}{dy} \gamma_s^{|S_i|} g_i \int \frac{d^3p}{(2\pi)^3} n_i(p;T,\mu_B,\mu_s) ,
\ee
where
\be 
n_i(p;T,\mu_B,\mu_s) = \left( e^{-(\sqrt{p^2+m_i^2} - \mu_B B_i - \mu_s S_i)/T} \pm 1 \right)^{-1}
\ee
is the Fermi (or Bose) distribution for given $T$, $\mu_B$, and strangeness chemical potential, $\mu_s$. Here, $p$ is 3-momentum,  $m_i$ is the mass of $i$-type particle, $g_i$, $B_i$ and $S_i$ are quantum degeneracy, baryon number and strangeness, respectively, and $\gamma_s$ is the strange quark fugacity. Some analyses fit the absolute yields, some only ratios of yields, which are independent of the freeze-out volume $dV/dy$. The fits typically include data for $\pi^+, K^+, p, \Lambda, \Xi$ and their antiparticles; some also include data for $\Omega$ and $\phi$ particles. 

The so-called ``hadrochemical horizon'' is not a sharp line. Hadrons undergo chemical reactions during the early stage of expansion following hadronization of the quark-gluon plasma until they ``freeze out'' after their last identity changing reaction. This process cannot be observed in action---at least we do not know how to experimentally track it---but it can be studied in detail in microscopic models of the evolution of the fireball after hadronization. Such a study was recently published by Reichert {et al.} \cite{Reichert:2020yhx} who used the Ultra-relativistic Quantum Molecular Dynamics (UrQMD) model to record the last moment of production of the emitted pions and correlate them with the coarse-grained thermodynamic variables describing the local environment during that moment. These pions continue to rescatter after their final production, which means that their {\em kinetic} freeze-out occurs later. 

The study has two limitations. First, whether the last moment of pion production is a good proxy for the hadrochemical horizon is open to debate. Most pions are emitted by resonance decay without changing the heavier flavor composition of the hadron. When the hadrochemical ratios are analyzed, all states that decay into each other by strong interactions are lumped together, which means that chemical freeze-out as operationally defined by the analysis of stable hadron ratios generally occurs earlier than the last moment of pion emission. Local temperatures recorded in this way therefore represent a lower bound  on the temperature of the hadrochemical horizon. The second limitation is that Ref.~\cite{Reichert:2020yhx} uses the UrQMD model to describe the collision from beginning to end without explicit creation of a deconfined phase. This means that some of the local temperatures at the last moment of pion production lie above the hadronic phase boundary and introduces an additional upward bias on the recorded temperature. The two countervailing effects both grow in size with increasing $\mu_B$.

With this in mind, it is still worth comparing the results of \cite{Reichert:2020yhx} with the STAR data, as shown in Figure~\ref{fig:Chem}. In a hybrid collision model, that includes a hydrodynamic expansion in the quark-gluon plasma phase, the average temperatures would shift to lower values, perhaps by one-third or half the width of the recorded temperature distribution (see Figure~3  in \cite{Reichert:2020yhx}) or 20--30 $\mathrm{MeV}$. Since the first limitation mentioned above generates a bias in the opposite direction (to lower temperatures) the real location of the hadrochemical freeze-out line as defined by stable hadron ratios is likely to lie 5--15~$\mathrm{MeV}$ below the black dots. A modified study within a hybrid collision model would be of considerable interest.

A cross-check of the thermodynamic freeze-out parameters at a given collision energy is possible by comparing certain volume-independent ratios of net quantum number fluctuations in the event ensemble with the corresponding ratio of susceptibilities that can be calculated on the lattice. This comparison has been made for net electric 
charge, $Q$, and net baryon number,  $B$, \cite{Bazavov:2012vg,Borsanyi:2013hza}. The susceptibilities, $\chi_n$, are defined as derivatives of the thermodynamic {pressure,} $P$, with respect to the chemical potential, $\mu_\alpha$, that is associated with the conserved quantity $\alpha=Q, B$:
\begin{equation}
\chi_n^{(\alpha)}(T,\mu_\alpha) = \frac{\partial^n(P/T^4)}{\partial(\mu_\alpha/T)^n} .
\end{equation}

The commonly used fluctuation measures (cumulants) are the mean, $M_\alpha$, the variance, $\sigma_\alpha^2$ 
and the   skewness, $S_\alpha$. They are related to the susceptibilities as
\begin{eqnarray}
& M_\alpha = V\chi_\alpha^{(1)},\quad 
\sigma_\alpha^2 = V\chi_\alpha^{(2)},\\
& S_\alpha = V^{-1/2}\chi_\alpha^{(3)}/(\chi_\alpha^{(2)})^{3/2}.
\end{eqnarray}
The unknown volume cancels in the ratios: 
\begin{equation}
R_{12} = M/\sigma^2,\quad R_{31} = S\sigma^{3/2}/M ,
\end{equation}
where the index $\alpha$, indicating which conserved quantum is being considered, is dropped for brevity.  

The experimental data for these ratios of cumulants generally agree with those derived from lattice-QCD calculations for the parameters deduced from the chemical freeze-out fits. There are quite a few caveats that come with such comparisons: Experiments measure fluctuations in momentum space, the lattice calculates fluctuations in position space; experiments average over a range of conditions under which particles are emitted, lattice simulations are for precisely fixed thermodynamic conditions; experiments cannot unambiguously separate initial state fluctuations from final-state (thermal) ones, lattice simulations only consider thermal fluctuations. (see \cite{Ratti:2021ubw} for a detailed discussion of this method and comparison with experimental data.) It is thus not entirely surpring that a recent analysis of net-kaon fluctuations using the $R_{12}$ measure seem to indicate that the strangeness content of the final hadron yields may freeze out at somewhat ($\approx$10 MeV) higher temperature \cite{Bellwied:2018tkc}. This could be attributed to their larger mass, which may let strange quarks lose their mobility slightly earlier when the QCD phase boundary is approached from above during the expansion of the fireball.

Overall, the experimental results are consistent with the principle that the fluctuations of conserved quantum numbers are frozen in at the moment when chemical reactions among hadrons freeze out. Since the quantum numbers probed by this method are carried by valence quarks, this observation is consistent with the idea that chemical freeze-out is related to 
valence-quark  
percolation. It is worth mentioning that the analyses for $B$ and $Q$ agree surprisingly well (see \cite{Borsanyi:2013hza}), because electric charge can be transported by pions, which diffuse easily and could be able to modify net charge fluctuations until kinetic freeze-out at lower temperature.

\section{Excluded Volume Bootstrap Model}
\label{sect4} 

As a schematic model of valence-quark  percolation, let us investigate the statistical bootstrap model with excluded volume of Hagedorn and Rafelski \cite{Hagedorn:1980kb}. The model is defined by the partition {function,} 
\be
Z(\beta,\mu,V) = \sum_{b=-\infty}^{\infty} e^{b\beta\mu} \int d^4p\, e^{-\beta\cdot p} \sigma(p,b,V) ,
\label{eq:Z}
\ee
where $b$ is the baryon number of the hadron and $\sigma(\beta,b,V)$ denotes the level density for hadrons with a given baryon number $b$. $\beta=1/T$ denotes the inverse temperature, $\mu$ are the chemical potentials for each quark flavor and $b$ counts the baryon number for each quark flavor. Here, $\mu_u = \mu_d = \mu_B/3$ and $\mu_s = 0$ are assumed.  The level density can be expressed as invariant phase space integral over the mass {spectrum} 
$\tau(m^2,b)$:
\ba
\sigma(p,b,V) &=& \delta^4(p)\delta_{b,0} +
\nn \\
& & \sum_{N=1}^{\infty} \frac{1}{N!}  \delta^4\left(p - \sum_i p_i \right) \sum_{\{b_i\}} \delta_{b,\sum_i b_i}
\nn \\
& & \times \prod_{i=1}^{N} \int d^4p_i\, \frac{2\Delta\cdot p}{(2\pi)^3} \tau(p_i^2,b_i) ,
\label{eq:LD}
\ea
where the first term corresponds to the vacuum state. Here, $\delta^4(\cdots)$ is the Dirac delta function, and $\delta_{x,y}$ is the Kronecker delta.  The $N$-th term involves a sum over partitions of the baryon number $b$ on $N$ hadron clusters with masses $m_i^2 = p_i^2$. The four-vector $\Delta^\mu \equiv \Delta u^\mu$, where  $\Delta$ is the unoccupied volume,  denotes the volume remaining after excluding the proper volumes,  $V_{{\rm cl},i}$, of all hadron clusters from the total fireball volume: $\Delta^\mu = V^\mu - \sum_i V_{{\rm cl},i} ^\mu$.

Following Hagedorn and Rafelski \cite{Hagedorn:1980kb}, the requirement that any hadron cluster is composed of smaller hadron clusters  is expressed by the {bootstrap}  condition  \cite{Hagedorn:1980kb},
\be
\sigma(p,b,V_{\rm cl}) 
= H\,\tau(p^2,b) ,
\label{eq:BS}
\ee
with the bootstrap constant $H = 0.724~\mathrm{(GeV)}^{-2}$. The bootstrap condition allows to generate the entire hadron cluster spectrum from a small number of ``elementary'' hadrons.  In \cite{Hagedorn:1980kb} only pions and nucleons are considered as input into the bootstrap equation; here, the entire set of quark model ground states is considered:  the pseudoscalar and vector meson nonets and the baryon and antibaryon octets and decuplets. Counting spin and isospin degrees of freedom this comprises 148 ``elementary'' states, which are labeledby the index $\alpha$. 

Inserting these states into Equation  (\ref{eq:Z}) and applying the Boltzmann approximation leads to the input partition function: 
\ba
\varphi(\beta,\mu) &=& \sum_b  e^{b\mu} \int d^4p\, e^{-\beta\cdot p}\, H \sum_\alpha g_\alpha \delta_+(p^2-m_\alpha^2) 
\nn \\
&=& 2\pi H T \sum_\alpha g_\alpha e^{b_\alpha\mu} m_\alpha K_1(m_\alpha/T) ,
\label{eq:phi}
\ea
where $g_\alpha$ denotes the degeneracy of each state and $b_\alpha$ its baryon number counting only $u$ and $d$ valence quarks. For values of $\mu_B$ approaching the nucleon mass, Fermi--Dirac quantum corrections must be taken into account for nucleons, which was done in the plots shown below.

The complete single-cluster partition function, $\phi(\beta,\mu)$, generated by Equation (\ref{eq:BS}), is related to $\varphi$ by the implicit equation,
\be
\varphi(\beta,\mu) = 2\phi(\beta,\mu) - e^{\phi(\beta,\mu)} +1 .
\label{eq:Phi}
\ee
The full partition function is obtained from $\phi$ as
\be
\ln Z(\beta,\mu,V) = -\frac{2\Delta}{(2\pi)^3H}\, \frac{\partial\phi(\beta,\mu)}{\partial\beta} .
\ee

In order to determine the unoccupied volume, $\Delta$, for a given hadron configuration within the fireball volume $V$, one needs to specify the proper volume of each cluster. Here, Ref.~\cite{Hagedorn:1980kb} 
is followed by using the MIT-Bag model relation $V_{\rm cl} = m_{\rm cl}/(4{\cal B})$, 
where $m_{\rm cl}$ is the cluster mass. This leads to the following equation for the unoccupied volume fraction \cite{Hagedorn:1980kb}:
\be
1-\xi \equiv \frac{\Delta}{V} = \left( 1 + \frac{\partial^2\phi/\partial\beta^2}{2(2\pi)^3H{\cal B}} \right)^{-1} .
\ee
Here the occupied (excluded) volume fraction, $\xi$, is introduced which is the parameter that controls percolation.

\begin{figure}[htb]
	\includegraphics[width=0.95\linewidth]{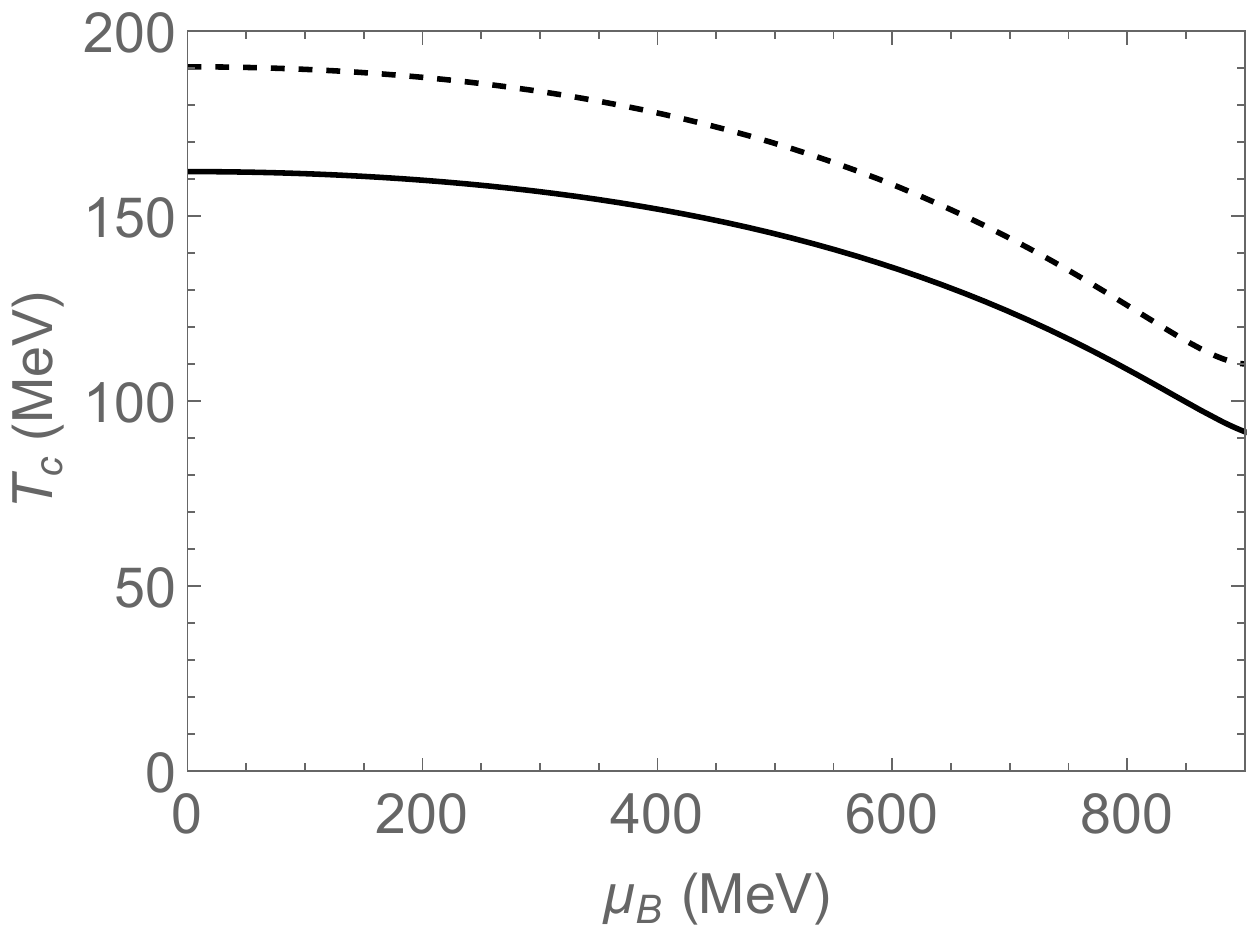}
	\caption{Critical temperature line, $T_c(\mu_B)$, delineating the boundary of hadronic matter in the excluded volume 	statistical bootstrap model. The solid line shows the boundary location when all quark model ground states are used 	at input into the statistical 	bootstrap; the dashed line shows the boundary when only pions and nucleons are 			considered as input.}
\label{fig:Tc}
\end{figure}

As shown in 
Ref.~\cite{Hagedorn:1980kb}, 
$\partial^2\phi/\partial\beta^2$ diverges when $\phi(\beta,\mu) = \ln 2$, and thus $\Delta$ 
vanishes: The entire fireball volume is occupied by a single hadronic cluster, which corresponds to the quark-gluon plasma. The line $\Delta 
= 0$ in the 
$T$--$\mu_B$  
diagram delineates the boundary of the hadronic phase in the statistical bootstrap model. 
 This line is denoted by $T_c(\mu_B)$ here, as shown in 
Figure~\ref{fig:Tc}. 
otherwise.
The solid line in Figure~\ref{fig:Tc} shows the phase boundary for the full set of ground state hadrons; for comparison, the dashed line shows the boundary when only pions and nucleons are considered as in 
Ref.~\cite{Hagedorn:1980kb}. 
The critical temperature, $T_c$,  
 at $\mu_B = 0$ for the full set of ground state hadrons is $T_c(0) = 162~\mathrm{MeV}$, in remarkably close coincidence with the pseudocritical temperature for the chiral phase transition found in lattice-QCD. 

\begin{figure}[htb]
	\includegraphics[width=0.95\linewidth]{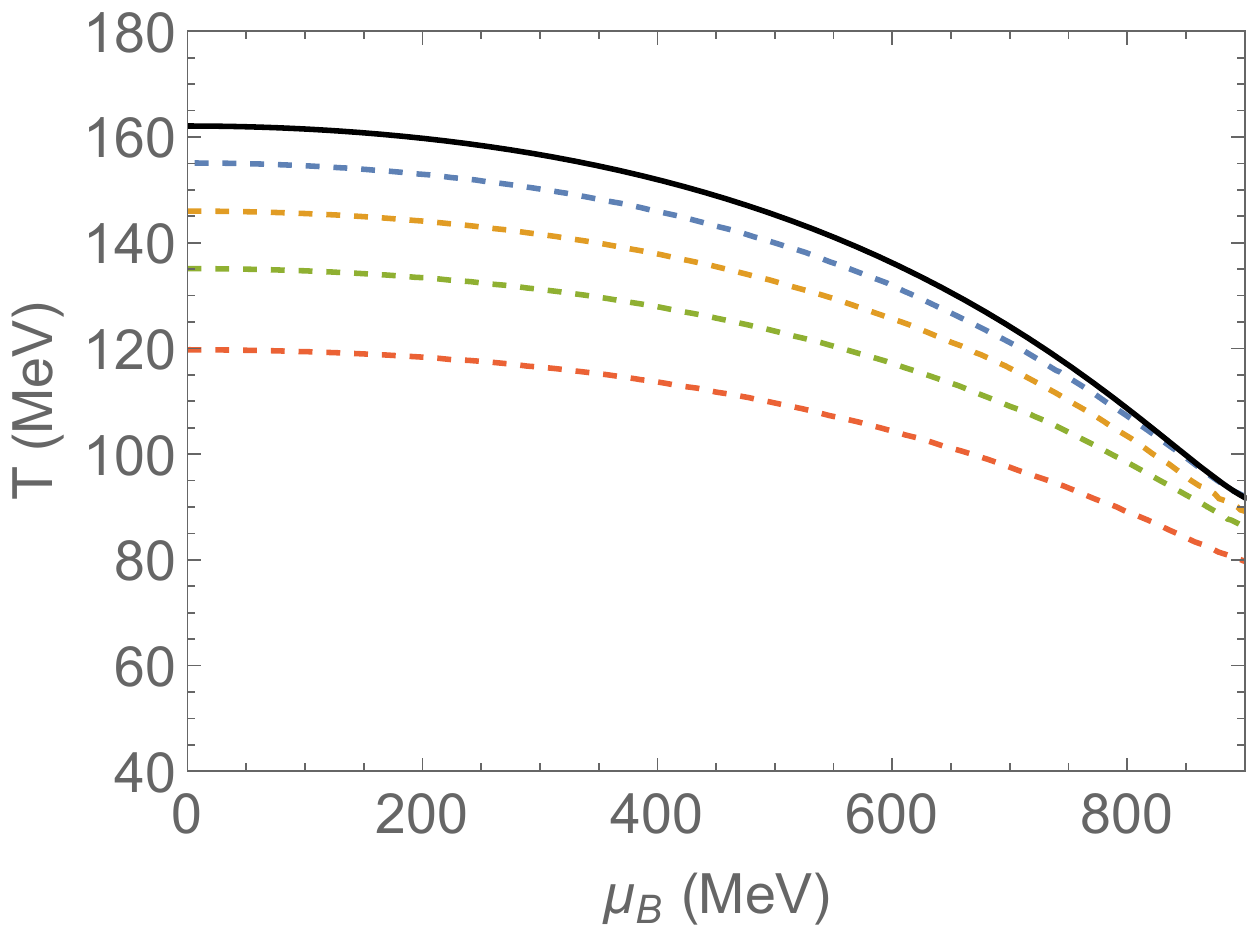} 
	\caption{Iso-occupancy contour lines, $T_\xi(\mu_B,\xi)$,  for hadronic matter in the statistical bootstrap model with 
	excluded volume. The dashed lines show the contours for the occupation fractions $\xi = 0.8, 0.6, 0.4,$ and $0.2$  		(top to bottom). The solid black line corresponds to $T_\xi(\mu_b,1) = T_c(\mu_B)$, also shown in  	
	Figure~\ref{fig:Chem}. The hadrochemical horizon identified by the valence-quark  percolation threshold is expected 
	to lie between the orange ($\xi=0.6$) and blue ($\xi=0.8$) dashed lines.}
\label{fig:Horizon}
\end{figure}

The line $T_c(\mu_B)$ traces the location of singularities in the partition function of the statistical bootstrap model with excluded volume; it delineates the thermodynamic phase boundary. Percolating clusters of hadrons already exist at lower temperatures corresponding to an unoccupied volume fraction $\Delta/V > 0$. Since the exact value of the percolation threshold, $\xi_c = 1-\Delta_c/V$, is not known, several contour lines of equal occupied (excluded) volume  fraction $\xi$ are shown in Figure~\ref{fig:Horizon}.  We denote these lines by $T_\xi(\mu_B,\xi)$ where obviously $T_\xi(\mu_B,1) = T_c(\mu_B)$. According to the discussion in Section  \ref{sect2}, the line $T_\xi(\mu_B,\xi_c)$ indicates the hadrochemical horizon, $T_h(\mu_B)$. Lacking precise knowledge of the dynamics of valence-quark  percolation,the precise location of this horizon is uncertain, but it  can be estimated that $T_h(\mu_B)$ lies between  $T_\xi(\mu_B,0.6)$ (orange line) 
and $T_\xi(\mu_B,0.8)$ (blue line).

\section{Summary and Outlook}

The chemical freeze-out line or chemical horizon, $T_\mathrm{chem}(\mu_B)$, of temperature vs. baryochemical potential,  serves as an experimentally measurable proxy for the phase boundary between matter composed of hadrons and quark matter. It is controlled by hadrochemical kinetics and, thus, somewhat sensitive to the overall size and expansion rate of the fireball. As chemical reactions among hadrons proceed primarily by quark (flavor) exchange, the chemical freeze-out line is expected to lie close to the valence-quark percolation line in the QCD phase diagram. 

In the low net-baryon density region of the phase diagram the chemical horizon agrees within experimental uncertainties with the pseudocritical thermodynamic phase boundary, $T_c(\mu_B)$, as determined by lattice QCD. At high net-baryon density the data indicate that the chemical horizon lies increasingly below the phase boundary. This could be caused by either a slower expansion rate during the time when the matter is near the phase boundary, or by changing chemical reaction kinetics in the baryon-dominated regime, or both. The origin of the empirical rule---``Cleymans' Law''---that chemical freeze-out occurs along a line of constant energy per particle \cite{Cleymans:1998fq} over the entire measured range is still unknown. {Some reminiscences of the late Jean Cleymans, who made fundamental contributions to the theory of relativistic heavy-ion collisions, are presented in Appendix}~\ref{appA}.  

It would be interesting to extend the theoretical simulations of the chemical freeze-out line in hadronic transport models to hybrid models of the reaction that include a hydrodynamic quark-gluon plasma phase. It would also be interesting to construct a more detailed model of valence-quark percolation and identify the precise location of the critical percolation line. Finally, it would be interesting to construct and study observables for valence-quark percolation that can be studied on the lattice.\\

\noindent
{\bf Note Added in Proof:} After submission of the manuscript, I became aware of a publication by Fukushima, Kojo and 
Weise~\cite{noteadded}, in which a similar concept of valence-quark percolation through meson exchange is discussed for the transition from nuclear matter to quark matter at high baryon density.

\vspace{6pt}

{\em Acknowledgments}:
This work was supported by the Office of Science of the U. S. Department of Energy (DOE)
under Grant DE-FG02-05ER41367. The author acknowledges support from Yale University for a sabbatical visit.

\appendix
\section{{Reminiscences of Jean Cleymans}} 
\label{appA}

I met Jean many times at conferences and always enjoyed the discussions with him. He was exemplary in being full of deep insights into the physics of relativistic heavy ion collisions and implications of experimental data and simultaneously gentle and modest. I also had multiple opportunities to interact with him during visits to Cape Town, which he had made  his home in the mid-1980s. That was a time when it required personal courage to ignore world-wide calls for boycotts of South Africa, but Jean was convinced that he could work for the benefit of South African society by helping to educate the future leaders, who would have to assume influential positions in science and education once apartheid ended; and end it did in 1991, sooner than many expected. 

Not long afterwards, I visited the University of Cape Town at Jean's invitation after a conference \cite{Stoecker:1997mu} that was organized by Horst St\"ocker in honor of Walter Greiner's $60^\textrm{th}$ birthday in Wilderness on the Indian Ocean coast of South Africa. During my visit, Jean arranged an excursion by boat for a few of us to Robben Island, the small island south of Cape Town where Nelson Mandela and other leaders of the African National Congress had been incarcerated from 1964 until 1982. At that time the site was not yet open to the public, and our small group had a personal tour of the island, the prison complex and the quarry in which Mandela and their fellow inmates had to labor under difficult conditions. It was an almost unreal experience to imagine the violence that was perpetrated at this site juxtaposed with its beautiful and tranquil nature, which included a small colony of penguins that nested at the shore. The visit gave us a sense of the incredible moral strength that allowed Mandela to maintain their conviction that reconciliation was possible after the end of apartheid through almost three decades of imprisonment. The {\em Truth and Reconciliation} process started when Mandela became President in 1994 and was in full swing during the time of our visit. 

My last visit to Cape Town at Jean's invitation occurred in 2017 on the occasion of a trip to South Africa with a small delegation from Brookhaven National Laboratory (BNL). The purpose of our trip was to explore possible collaborations between BNL and various South African institutions, one of which was the University of Cape Town. Jean was as gracious host as ever, and he arranged for a dinner at the residence of the President of the University. It was clear that the University, which always had been open to students of color, had become an even more integrated institution of higher learning, which served its role of educating future leaders of the country that Jean had envisioned well. A younger generation of scientists working on relativistic heavy-ion physics is now carrying on his vision. 

Openness and inclusivity often do not make life easier in the short term, but ultimately it is the only way by which societies can make lasting progress. Seeing this and working towards it with an eye to the long term was one of Jean's strengths, which is also reflected in the way he approached science. With this he served as a role model for many of his friends, younger colleagues, and students. It is his legacy. 



\begin{thebibliography}{999}

\bibitem[Alford \em{et~al.}(2008)Alford, Schmitt, Rajagopal, and
  Sch\"afer]{Alford:2007xm}
Alford, M.G.; Schmitt, A.; Rajagopal, K.; Sch\"afer, T.
\newblock {Color superconductivity in dense quark matter}.
\newblock {\em Rev. Mod. Phys.} {\bf 2008}, {\em 80},~1455--1515.

\bibitem[Andronic \em{et~al.}(2010)Andronic et~al.]{Andronic:2009gj}
Andronic, A.; Blaschke, D.; Braun-Munzinger, P.; Cleymans, J.; Fukushima, K.; McLerran, L.D.; ; Oeschler, H.; Pisarski, R.D.; Redlich, K.; Stachel, J.; et al. 
\newblock {Hadron Production in Ultra-relativistic Nuclear Collisions:
  Quarkyonic Matter and a Triple Point in the Phase Diagram of QCD}.
\newblock {\em Nucl. Phys. A} {\bf 2010}, {\em 837},~65--86.

\bibitem[Hagedorn and Rafelski(1980)]{Hagedorn:1980kb}
Hagedorn, R.; Rafelski, J.
\newblock {Hot hadronic matter and nuclear collisions}.
\newblock {\em Phys. Lett. B} {\bf 1980}, {\em 97},~136.

\bibitem[Bazavov \em{et~al.}(2017)Bazavov et~al.]{Bazavov:2017dus}
Bazavov, A.; Ding, H.T.; Hegde, P.; Kaczmarek, O.; Karsch, F.; Laermann, E.;  Maezawa, Y.; Mukherjee, S.; Ohno, H.; Wagner, M.; et al.
\newblock {The QCD equation of state to $\mathcal{O}(\mu_B^6)$ from lattice
  QCD}.
\newblock {\em Phys. Rev. D} {\bf 2017}, {\em 95},~054504.

\bibitem[Borsanyi \em{et~al.}(2020)Borsanyi, Fodor, Guenther, Kara, Katz,
  Parotto, Pasztor, Ratti, and Szabo]{Borsanyi:2020fev}
Borsanyi, S.; Fodor, Z.; Guenther, J.N.; Kara, R.; Katz, S.D.; Parotto, P.;
  Pasztor, A.; Ratti, C.; Szabo, K.K.
\newblock {QCD Crossover at Finite Chemical Potential from Lattice
  Simulations}.
\newblock {\em Phys. Rev. Lett.} {\bf 2020}, {\em 125},~052001.

\bibitem[Hatsuda \em{et~al.}(2006)Hatsuda, Tachibana, Yamamoto, and
  Baym]{Hatsuda:2006ps}
Hatsuda, T.; Tachibana, M.; Yamamoto, N.; Baym, G.
\newblock {New critical point induced by the axial anomaly in dense QCD}.
\newblock {\em Phys. Rev. Lett.} {\bf 2006}, {\em 97},~122001.

\bibitem[Bors\'anyi \em{et~al.}(2021)Bors\'anyi, Fodor, Guenther, Kara, Katz,
  Parotto, P\'asztor, Ratti, and Szab\'o]{Borsanyi:2021sxv}
Bors\'anyi, S.; Fodor, Z.; Guenther, J.N.; Kara, R.; Katz, S.D.; Parotto, P.;
  P\'asztor, A.; Ratti, C.; Szab\'o, K.K.
\newblock {Lattice QCD equation of state at finite chemical potential from an
  alternative expansion scheme}.
\newblock {\em Phys. Rev. Lett.} {\bf 2021}, {\em 126},~232001.

\bibitem[Fu \em{et~al.}(2020)Fu, Pawlowski, and Rennecke]{Fu:2019hdw}
Fu, W.j.; Pawlowski, J.M.; Rennecke, F.
\newblock {QCD phase structure at finite temperature and density}.
\newblock {\em Phys. Rev. D} {\bf 2020}, {\em 101},~054032.

\bibitem[Critelli \em{et~al.}(2017)Critelli, Noronha, Noronha-Hostler,
  Portillo, Ratti, and Rougemont]{Critelli:2017oub}
Critelli, R.; Noronha, J.; Noronha-Hostler, J.; Portillo, I.; Ratti, C.;
  Rougemont, R.
\newblock {Critical point in the phase diagram of primordial quark-gluon matter
  from black hole physics}.
\newblock {\em Phys. Rev. D} {\bf 2017}, {\em 96},~096026.

\bibitem[Bazavov \em{et~al.}(2019)Bazavov et~al.]{HotQCD:2018pds}
Bazavov, A.; Ding, H.T.; Hegde, P.; Kaczmarek, O.; Karsch, F.; Karthik, N.;  Laermann, E.; Lahiri, A.; Larsen, R.;  Steinbrecher, P.; et al.
\newblock {Chiral crossover in QCD at zero and non-zero chemical potentials}.
\newblock {\em Phys. Lett. B} {\bf 2019}, {\em 795},~15--21.

\bibitem[Magas and Satz(2003)]{Magas:2003wi}
Magas, V.; Satz, H.
\newblock {Conditions for confinement and freezeout}.
\newblock {\em Eur. Phys. J. C} {\bf 2003}, {\em 32},~115--119.

\bibitem[Armesto \em{et~al.}(1996)Armesto, Braun, Ferreiro, and
  Pajares]{Armesto:1996kt}
Armesto, N.; Braun, M.A.; Ferreiro, E.G.; Pajares, C.
\newblock {Percolation approach to quark-gluon plasma and J/psi
  suppression}.
\newblock {\em Phys. Rev. Lett.} {\bf 1996}, {\em 77},~3736--3738.

\bibitem[Braun \em{et~al.}(1999)Braun, Pajares, and Ranft]{Braun:1997ch}
Braun, M.A.; Pajares, C.; Ranft, J.
\newblock {Fusion of strings versus percolation and the transition to the quark
  gluon plasma}.
\newblock {\em Int. J. Mod. Phys. A} {\bf 1999}, {\em 14},~2689--2704.

\bibitem[Kovner \em{et~al.}(1995)Kovner, McLerran, and Weigert]{Kovner:1995ja}
Kovner, A.; McLerran, L.D.; Weigert, H.
\newblock {Gluon production from nonAbelian Weizsacker-Williams fields in
  nucleus-nucleus collisions}.
\newblock {\em Phys. Rev. D} {\bf 1995}, {\em 52},~6231--6237.

\bibitem[McLerran(2007)]{McLerran:2007zzc}
McLerran, L.
\newblock {Color glass condensate and glasma}.
\newblock {\em AIP Conf. Proc.} {\bf 2007}, {\em 917},~219--230.

\bibitem[Chodos \em{et~al.}(1974)Chodos, Jaffe, Johnson, and
  Thorn]{Chodos:1974pn}
Chodos, A.; Jaffe, R.L.; Johnson, K.; Thorn, C.B.
\newblock {Baryon structure in the bag theory}.
\newblock {\em Phys. Rev. D} {\bf 1974}, {\em 10},~2599.

\bibitem[Thomas \em{et~al.}(1981)Thomas, Theberge, and Miller]{Thomas:1981vc}
Thomas, A.W.; Theberge, S.; Miller, G.A.
\newblock {Cloudy bag model of the nucleon}.
\newblock {\em Phys. Rev. D} {\bf 1981}, {\em 24},~216.

\bibitem[Stauffer and Aharony(2018)]{Stauffer:2018}
Stauffer, D.; Aharony, A.
\newblock {\em Introduction to Percolation Theory};
\& Francis: London, UK, 1994.  

\bibitem[Grimaldi(2017)]{Grimaldi:2007}
Grimaldi, C.
\newblock Tree-ansatz percolation of hard spheres.
\newblock {\em  J. Chem. Phys.} {\bf 2017}, {\em 147},~074502.

\bibitem[Desmond and Weeks(2014)]{Desmond:2014}
Desmond, K.W.; Weeks, E.R.
\newblock Influence of particle size distribution on random close packing of
  spheres.
\newblock {\em Phys. Rev. E} {\bf 2014}, {\em 90},~022204.

\bibitem[Cleymans and Satz(1993)]{Cleymans:1992zc}
Cleymans, J.; Satz, H.
\newblock {Thermal hadron production in high-energy heavy ion collisions}.
\newblock {\em Z. Phys. C} {\bf 1993}, {\em 57},~135--148.

\bibitem[Suhonen \em{et~al.}(1993)Suhonen, Cleymans, Redlich, and
  Satz]{Suhonen:1993zy}
Suhonen, E.; Cleymans, J.; Redlich, K.; Satz, H.
\newblock {Hadron production ratios as probes of confinement and freezeout}.
\newblock \emph{arXiv}  \textbf{1993}, arXiv:hep-ph/9310345.


\bibitem[Letessier \em{et~al.}(1994)Letessier, Rafelski, and
  Tounsi]{Letessier:1994cn}
Letessier, J.; Rafelski, J.; Tounsi, A.
\newblock {Strangeness and particle freezeout in nuclear collisions at
 14.6~GeV $A$}. 
\newblock {\em Phys. Lett. B} {\bf 1994}, {\em 328},~499--505.

\bibitem[Koch \em{et~al.}(1986{\natexlab{a}})Koch, M\"uller, and
  Rafelski]{Koch:1986hf}
Koch, P.; M\"uller, B.; Rafelski, J.
\newblock {Strangeness production and evolution in quark gluon plasma}.
\newblock {\em Z. Phys. A} {\bf 1986}, {\em 324},~453--463.
\bibitem[Koch \em{et~al.}(1986{\natexlab{b}})Koch, M\"uller, and
  Rafelski]{Koch:1986ud}
Koch, P.; M\"uller, B.; Rafelski, J.
\newblock {Strangeness in relativistic heavy ion collisions}.
\newblock {\em Phys. Rept.} {\bf 1986}, {\em 142},~167--262.

\bibitem[Braun-Munzinger \em{et~al.}(2004)Braun-Munzinger, Stachel, and
  Wetterich]{Braun-Munzinger:2003htr}
Braun-Munzinger, P.; Stachel, J.; Wetterich, C.
\newblock {Chemical freezeout and the QCD phase transition temperature}.
\newblock {\em Phys. Lett. B} {\bf 2004}, {\em 596},~61--69.

\bibitem[Harris and M\"uller(1996)]{Harris:1996zx}
Harris, J.W.; M\"uller, B.
\newblock {The Search for the quark-gluon plasma}.
\newblock {\em Ann. Rev. Nucl. Part. Sci.} {\bf 1996}, {\em 46},~71--107.

\bibitem[Cleymans and Redlich(1998)]{Cleymans:1998fq}
Cleymans, J.; Redlich, K.
\newblock {Unified description of freezeout parameters in relativistic heavy
  ion collisions}.
\newblock {\em Phys. Rev. Lett.} {\bf 1998}, {\em 81},~5284--5286.

\bibitem[Borsanyi \em{et~al.}(2010)Borsanyi, Endrodi, Fodor, Jakovac, Katz,
  Krieg, Ratti, and Szabo]{Borsanyi:2010cj}
Borsanyi, S.; Endrodi, G.; Fodor, Z.; Jakovac, A.; Katz, S.D.; Krieg, S.;
  Ratti, C.; Szabo, K.K.
\newblock {The QCD equation of state with dynamical quarks}.
\newblock {\em J. High Energy Phys.} {\bf 2010}, {\em 11},~077.

\bibitem[Adamczyk \em{et~al.}(2017)Adamczyk et~al.]{STAR:2017sal}
 Adamczyk, L.; Adkins, J.K.; Agakishiev, G.; Aggarwal, M.M.; Ahammed, Z.; Ajitanand, N.N.; Alekseev, I.; Anderson, D.M.; Aoyama, R.; Aparin, A.; et al.
 (STAR Collaboration).
\newblock {Bulk properties of the medium produced in relativistic heavy-ion
  collisions from the beam energy scan program}.
\newblock {\em Phys. Rev. C} {\bf 2017}, {\em 96},~044904.

\bibitem[Becattini \em{et~al.}(2006)Becattini, Manninen, and
  Gazdzicki]{Becattini:2005xt}
Becattini, F.; Manninen, J.; Gazdzicki, M.
\newblock {Energy and system size dependence of chemical freeze-out in
  relativistic nuclear collisions}.
\newblock {\em Phys. Rev. C} {\bf 2006}, {\em 73},~044905.

\bibitem[Andronic \em{et~al.}(2006)Andronic, Braun-Munzinger, and
  Stachel]{Andronic:2005yp}
Andronic, A.; Braun-Munzinger, P.; Stachel, J.
\newblock {Hadron production in central nucleus-nucleus collisions at chemical
  freeze-out}.
\newblock {\em Nucl. Phys. A} {\bf 2006}, {\em 772},~167--199.

\bibitem[Flor \em{et~al.}(2021)Flor, Olinger, and Bellwied]{Flor:2020fdw}
Flor, F.A.; Olinger, G.; Bellwied, R.
\newblock {Flavour and energy dependence of chemical freeze-out temperatures in
  relativistic heavy ion collisions from RHIC-BES to LHC energies}.
\newblock {\em Phys. Lett. B} {\bf 2021}, {\em 814},~136098.

\bibitem[Reichert \em{et~al.}(2020)Reichert, Inghirami, and
  Bleicher]{Reichert:2020yhx}
Reichert, T.; Inghirami, G.; Bleicher, M.
\newblock {Probing chemical freeze-out criteria in relativistic nuclear
  collisions with coarse grained transport simulations}.
\newblock {\em Eur. Phys. J. A} {\bf 2020}, {\em 56},~267.

\bibitem[Becattini \em{et~al.}(2001)Becattini, Cleymans, Keranen, Suhonen and Redlich]{Becattini:2000jw}
Becattini, F.; Cleymans, J.; Keranen, A.; Suhonen, E.; Redlich, K.
\newblock {Features of particle multiplicities and strangeness production in central heavy ion collisions between 1.7$A$ 
and 158$A$ GeV/$c$}.
\newblock {\em Phys. Rev. C} {\bf 2001}, {\em 64},~024901.

\bibitem[Bazavov \em{et~al.}(2012)Bazavov et~al.]{Bazavov:2012vg}
Bazavov, A.; Ding, H.T.; Hegde, P.; Kaczmarek, O.; Karsch, F.; Laermann, E.;  Mukherjee, S.; Petreczky, P.; Schmidt, C.; 
Smith, D;  et al. 
\newblock {Freeze-out conditions in heavy ion collisions from QCD
  thermodynamics}.
\newblock {\em Phys. Rev. Lett.} {\bf 2012}, {\em 109},~192302.

\bibitem[Borsanyi \em{et~al.}(2013)Borsanyi, Fodor, Katz, Krieg, Ratti, and
  Szabo]{Borsanyi:2013hza}
Borsanyi, S.; Fodor, Z.; Katz, S.D.; Krieg, S.; Ratti, C.; Szabo, K.K.
\newblock {Freeze-out parameters: Lattice meets experiment}.
\newblock {\em Phys. Rev. Lett.} {\bf 2013}, {\em 111},~062005.

\bibitem[Ratti and Bellwied(2021)]{Ratti:2021ubw}
Ratti, C.; Bellwied, R.
\newblock {\em {The Deconfinement Transition of QCD. 
 Theory Meets Experiment}};
 Springer Nature Switzerland AG: Cham, Switzerland,  2021

\bibitem[Bellwied \em{et~al.}(2019)Bellwied, Noronha-Hostler, Parotto,
  Portillo~Vazquez, Ratti, and Stafford]{Bellwied:2018tkc}
Bellwied, R.; Noronha-Hostler, J.; Parotto, P.; Portillo~Vazquez, I.; Ratti,
  C.; Stafford, J.M.
\newblock {Freeze-out temperature from net-kaon fluctuations at energies
  available at the BNL Relativistic Heavy Ion Collider}.
\newblock {\em Phys. Rev. C} {\bf 2019}, {\em 99},~034912.

\bibitem{noteadded} 
Fukushima, K.; Kojo, T.;  Weise, W. 
Hard-core deconfinement and soft-surface delocalization from nuclear to quark matter. {\em Phys. Rev. D} {\bf 2020}, {\em 102},~096017. 

\bibitem[Stoecker \em{et~al.}(1997)Stoecker, Gallmann, and
  Hamilton]{Stoecker:1997mu}
Stoecker, H.; Gallmann, A.; Hamilton, J.H. (Eds.) 
\newblock {\em {Structure of Vacuum and Elementary Matter,  Proceedings of the 
  International Conference on Nuclear Physics at the Turn of the Millennium,
  Wilderness, South Africa, 10--16 March  1996}};  World Scientific Co.: Singapore, 1997. 

\end{thebibliography}
\end{document}